\documentclass[lettersize,journa]{IEEEtran}
\IEEEoverridecommandlockouts
\usepackage{cite}
\usepackage{amsmath,amssymb,amsfonts}
\usepackage{algorithmic}
\usepackage{graphicx}
\usepackage{textcomp}
\usepackage{xcolor}
\usepackage{url}
\usepackage{hyperref}
\usepackage{tikz}
\usepackage{blindtext}
\usepackage{mathptmx}
\usepackage{balance}
\usepackage[normalem]{ulem}

\makeatletter
\def\ps@IEEEtitlepagestyle{
  \def\@oddfoot{\mycopyrightnotice}
  \def\@evenfoot{}
}
\def\mycopyrightnotice{
  {\footnotesize
  \begin{minipage}{\textwidth}
  \centering
  Copyright~\copyright~2022 IEEE. Personal use of this material is permitted. Permission from IEEE must be obtained for all other uses, in any current or future media, including reprinting/republishing this material for advertising or promotional purposes, creating new collective works, for resale or redistribution\\to servers or lists, or reuse of any copyrighted component of this work in other works.
  \end{minipage}
  }
}

\begin{document}

\title{Space-Terrestrial Integrated Internet of Things: Challenges and Opportunities}

\newcommand{\jf}[1]{{\textcolor{orange}{#1}}}
\newcommand{\fv}[2]{{\textcolor{red}{#2}}}
\newcommand{\oio}[1]{{\textcolor{purple}{#1}}}
\newcommand{\new}[1]{{\textcolor{black}{#1}}}

\author{
\IEEEauthorblockN{
Juan A. Fraire\IEEEauthorrefmark{1}\IEEEauthorrefmark{2}\IEEEauthorrefmark{3},
Oana Iova\IEEEauthorrefmark{1},
Fabrice Valois\IEEEauthorrefmark{1}
\\
}
\IEEEauthorblockA{
\IEEEauthorrefmark{1}Univ Lyon, Inria, INSA Lyon, CITI, F-69621 Villeurbanne, France\\
\IEEEauthorrefmark{2}Saarland University, Saarland Informatics Campus E1\,3, 66123 Saarbrücken, Germany\\
\IEEEauthorrefmark{3}CONICET - Universidad Nacional de Córdoba, Argentina
}
}

\maketitle

\begin{abstract}
Large geographical regions of our planet remain uncovered by terrestrial network connections.
Sparse and dense constellations of near-Earth orbit satellites can bridge this gap by providing Internet of Things (IoT) connectivity on a world-wide scale in a flexible and cost-effective manner. 
This paper presents \new{STEREO:} a novel Space-Terrestrial Integrated IoT Architecture spanning direct- and indirect-to-satellite access from IoT assets on the surface.
Framed on the identified requirements, we analyze NB-IoT and LoRa/LoRaWAN features to put these technologies forward as appealing candidates for future satellite IoT deployments.
Finally, we list and discuss the key open research challenges to be addressed in order to achieve a successful space-terrestrial IoT integration. 
\end{abstract}



\section{Introduction}\label{sec:intro}

The recent spur in space projects~\cite{leyva2020leo} has revamped the interest in satellite communication.
This is especially observed in the Internet of Things (IoT) community that constantly seeks to diversify the application scenarios~\cite{centenaro2021survey} while providing network coverage anywhere in the world. 
The unique characteristics of satellites in the \textit{new space} context (cheap launch and quick procurement of inexpensive nano-satellites a.k.a. \textit{CubeSats}) enable architectural alternatives for IoT networks with degrees of scale and flexibility hitherto impossible~\cite{sweeting2018modern}.

\begin{figure}
	\centering
	\includegraphics[width=\linewidth]{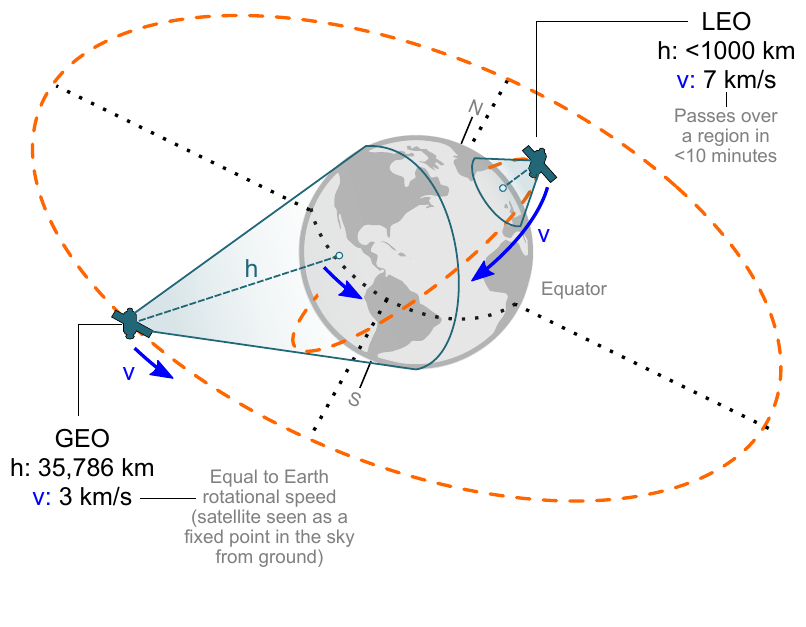}
	\caption{LEO and GEO orbit illustration.}
	\label{fig:orbits}
\end{figure}

Satellites deployed in Geosynchronous Orbits (GEO) exhibit a rotation period equal to that of the Earth (appearing motionless to an observer on the ground), which can offer continuous network connectivity over a specific area from 35,786~km height (Fig.~\ref{fig:orbits} and Table~\ref{tab:orbits}).
On the other hand, Low Earth Orbit (LEO) satellites move at $\sim$7~km/s, at lower altitudes (between 160 km and 1,000 km), and can provide intermittent and regular network connectivity at predictable time intervals. 
When deployed in constellations, LEO satellites can increase the revisit frequency, but at least 60 of them are needed to ensure continuous coverage. 

\begin{table}[]
\centering
\caption{LEO and GEO orbit characteristics}
\label{tab:orbits}
	\begin{tabular}{l|l|l|}
\cline{2-3}
                                        & LEO                   & GEO       \\ \hline
\multicolumn{1}{|l|}{Altitude}          & 160-1000 km           & 35786 km  \\ \hline
\multicolumn{1}{|l|}{Orbital Period}    & $\sim$90 minutes            & 24 hours  \\ \hline
\multicolumn{1}{|l|}{Pass duration}     & \textless{}10 minutes & Permanent \\ \hline
\multicolumn{1}{|l|}{Earth surface coverage}     & \textless{}1.5\% & $\sim$30\% \\ \hline
\multicolumn{1}{|l|}{Global coverage}   & 60+ needed            & 3+ needed \\ \hline
\multicolumn{1}{|l|}{Propagation delay} & 7 ms                  & 120 ms    \\ \hline
\multicolumn{1}{|l|}{Lifetime}          & \textless{}5 years    & 15 years  \\ \hline
\multicolumn{1}{|l|}{Doppler}           & Yes                   & No        \\ \hline
\end{tabular}
\end{table}

By embarking IoT equipment on board of these satellites, new connectivity opportunities arise.
The advancement in communication technologies makes it possible today to have a direct communication between IoT devices and satellites using the same technologies as terrestrial IoT networks~\cite{fraire2019direct}, which until recent years was unheard of. 
The most notable advances in such technologies involve LoRa/LoRaWAN~\cite{colavolpe2019reception} and NB-IoT~\cite{ouvry2018ultra}, which offer long range communication capabilities and a reduced device energy consumption (18 mA @7dBm). 

The opportunities brought about by a satellite-boosted global network for IoT are immeasurable.
\new{Existing terrestrial IoT networks could take advantage of the enhanced satellite coverage for rural and out-of-reach areas, offering ubiquitous global connectivity services (especially in regions where coverage is otherwise technically and/or economically not viable for existing IoT infrastructure), and enabling a plethora of new applications opportunities.}
These will impact several sectors such as:
(i) global asset tracking in land, sea and air, transport of vehicles, fleets, objects and materials;
(ii) monitoring of environmental parameters over geographical areas where terrestrial networks are not present (inaccessible forests, large deserts, oceans);
(iii) cross-border energy production management (in gas and oil extraction, renewable solar farms and offshore wind-parks).
Finally, smart metering, agriculture utilities, and effective healthcare solutions already available in urbanized regions could be brought to faraway lands with isolated individuals~\cite{yang2019federated}.

A massive data collection will thus be possible leveraging globally connected assets with the potential of creating the largest networked ecosystem where unparalleled interaction could take place.
The satellite IoT phenomena interestingly occurs in times where federated machine learning is starting to unblock secure, privacy-preserving and collaborative training approaches~\cite{yang2019federated}.
In this context, satellite IoT could contribute with the potential of collecting and sharing world-wide data to enable future intelligent applications such as driver-less cars, automated medical care, finance, and insurance. 

These opportunities come at the expense of overcoming non-trivial technical challenges due to the long space-terrestrial channel, the orbital dynamics of satellites, and the highly constrained IoT devices on the ground.
Existing IoT \textit{medium access control} schemes need to be revised and/or extended to scale up to thousands of IoT devices having to simultaneously communicate with the gateway in under 10 minutes (the pass duration of a LEO satellite). 
Crucial IoT \textit{core network functionalities} such as mobility and management need to be identified and virtualized 
and flexibly placed in space and/or ground elements as mandated by the time-evolving and disruptive nature of satellite IoT network topology. 

\new{This paper is the first to present a comprehensive vision of a \underline{S}pace-\underline{Ter}restrial Int\underline{e}grated I\underline{o}T we coined STEREO. The contribution  is the first to span network architecture, applicable terrestrial technologies and related challenges in porting them to the space domain. Specifically:}
\begin{enumerate}
    \item We introduce a reference \textbf{architecture} for the integration of space-terrestrial networks as a whole, leveraging the state-of-the art of space and terrestrial technologies, protocols and procedures. 
    \item We review and we propose the set of \textbf{technologies} to be used in the future space-terrestrial integrated networks, focusing on the satellite IoT communication.
    \item We highlight the main research \textbf{challenges} for realizing a world-wide IoT connectivity. 
\end{enumerate}
Furthermore, we discuss that unprecedented solutions arise from (i) exploiting scheduling solutions based on the predictable nature of orbital mechanics, (ii) leveraging the delay-tolerant nature of the IoT traffic, (iii) learning from frequent revisits to service areas by multiple passing-by satellites, and (iv) strategically deploying virtualized network functions in ground and orbit assets.

The remainder of this paper paper is organized as follows.
The envisioned satellite IoT architecture to accomplish the former integration is presented in Section~\ref{sec:taxonomy}.
The key technologies enabling satellite IoT communication are presented in Section~\ref{sec:background}.
The main challenges and open research topics are identified and formulated in Section~\ref{sec:challenges}, 
Finally, Section~\ref{sec:conclusion} presents the outlook of this work.

\section{The Future IoT Network Architecture}\label{sec:taxonomy}

\begin{figure*}[h!]
	\centering
	\includegraphics[width=\textwidth]{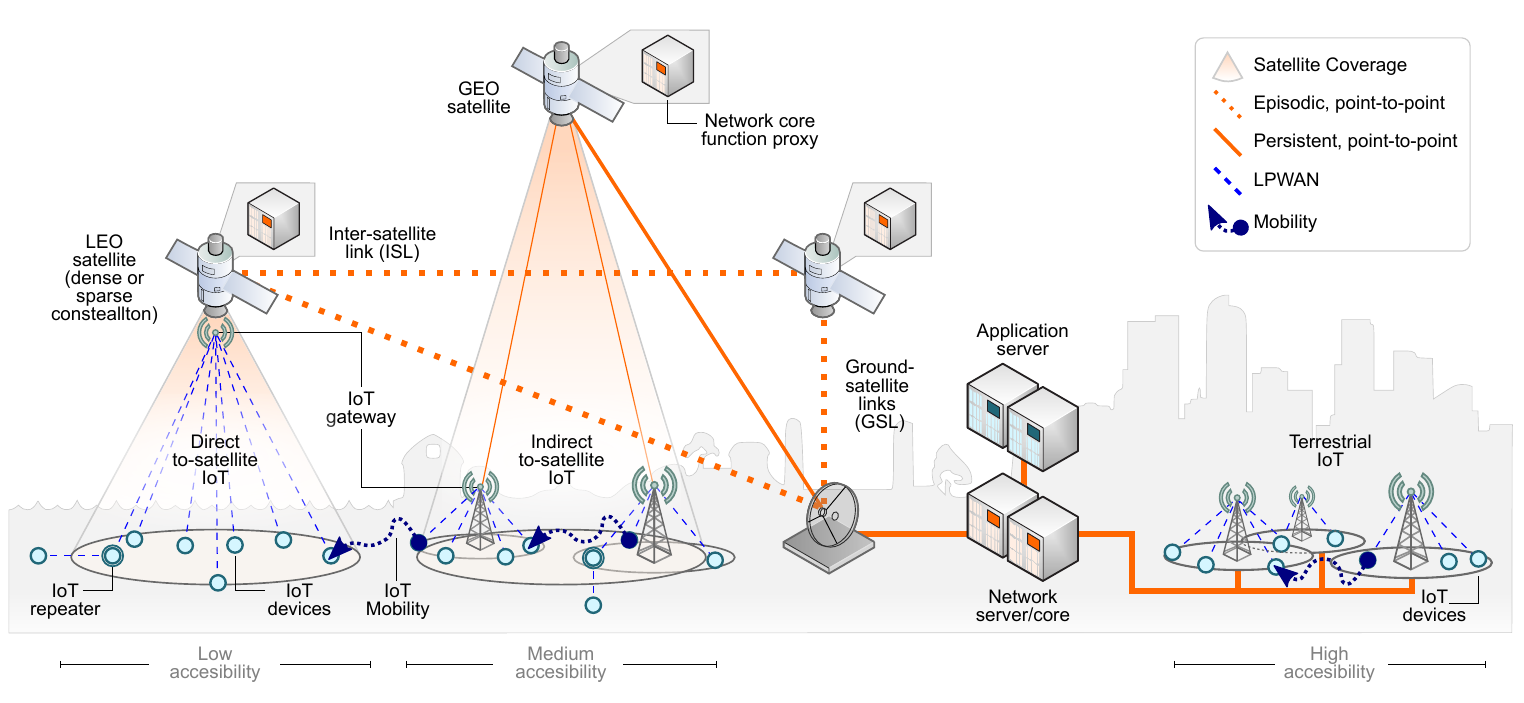}
	\caption{\new{STEREO} architecture.}
	\label{fig:architecture}
\end{figure*}

In the context of enabling access equity by bringing together satellite communication and IoT technologies, we propose the \new{STEREO} architecture, depicted in Fig~\ref{fig:architecture}. 
We claim that the future IoT networks have to take advantage of all available technologies by deploying and mixing both terrestrial and satellite networks. 
The decision of which specific technology to be deployed in a given area should be taken by accounting for the specificity of that geographical area, the availability of electrical power, and the density of end-devices. 
In this respect, besides the high accessibility in urban areas, we identified two possible deployment scenarios: \textit{indirect} and \textit{direct} satellite IoT.
It is important to not see these different deployments as individual, but as interconnected networks belonging to the same global network.


\subsection{Indirect to Satellite IoT} 
Outside highly accessible, dense, urban areas, we find rural areas with a high concentration of end-devices that justifies the deployment of dedicated ground IoT gateways to serve them.
However, the terrain and local conditions (i.e., lack of cellular coverage, impossibility to deploy fiber) might make it difficult to have an available infrastructure to transport data from the gateways to the core network. 
In this case, using satellites to serve as backhaul for these gateways placed on ground surface is a very appealing solution.
The stable position in the sky of of GEO satellites makes of them a perfect fit to relay data from ground gateways equipped with fixed high-gain antennas and a suitable power source (e.g., solar or electric grid) capable of establishing long-range links to geostationary orbit.

As end-devices do not reach the satellite directly (but via gateways) this type of deployment has been named indirect-to-satellite IoT communication (ItS-IoT). 
The protocols to be used can be based on a clearly separated ground and space domains. 
Specifically, terrestrial IoT (e.g., LoRa/LoRaWAN, NB-IoT) will continue to be used the same way between end-devices and gateways, with some adjustments to account for the higher delay between the gateway and the network server, as a consequence of the satellite links (e.g., adapting the times for scheduling downlink communication). 
Existing space-specific protocols and technologies (e.g., CCSDS-based protocols, discussed in the next section) can then be implemented on the gateway-to-satellite link.
The challenge in ItS-IoT is integrating space and terrestrial protocols into an efficient end-to-end \new{STEREO} architecture, as we detail in Section~\ref{sec:challenges}.

\subsection{Direct to Satellite IoT} 
Applications that need to function in less accessible regions (i.e., oceans, mountains, poles) might not justify or even hinder the deployment of IoT gateways on ground.
In such scenarios, IoT devices should rather directly access the satellite hosting an on-board IoT gateway.
As GEO links are not suitable due to the large range, LEO satellites emerge as the most appealing approach. 
Flying at less than 1000 km above Earth, the channel with LEO satellites can be set up to meet the margins required by terrestrial IoT protocols, even with low-cost antennas on the IoT device (and without any other modification to the device). 
\new{Recent in-orbit deployments including the \href{https://www.univ-grenoble-alpes.fr/english/latest-grenoble-made-nano-satellite-launched-into-space--987001.kjsp?RH=2320611992758370}{ThingSat}, \href{https://www.hackster.io/news/fossasat-1-an-open-source-satellite-for-the-internet-of-things-7f31cab00ef5}{FossaSat}, \href{https://space.skyrocket.de/doc_sdat/lacunasat-3.htm}{LacunaSat}, and \href{https://alen.space/alen-space-validates-in-orbit-the-payload-of-the-first-cubesat-of-sateliots-constellation/}{Sateliot} satellites proved the feasibility of the approach, also discussed in related research papers~\cite{colavolpe2019reception,ouvry2018ultra}}.
Moreover, modern and cost-effective LEO nano-satellite platforms can be leveraged to accommodate the reduced power and volume requirements of IoT gateways.

The main challenge with direct-to-satellite IoT (DtS-IoT) communication is that DtS-IoT enables a high-speed flying IoT gateway, with a highly varying channel, over a predictable orbital trajectory. 
More specifically, the duration of a typical satellite pass over a given region is in the order of 10 to 3 minutes for a perfectly zenithal pass and a pass over the horizon, respectively. 
Hence, during this period, the channel conditions vary drastically from more than 2,000 km to the actual satellite altitude (at the zenith position as seen from the device perspective).
Since the coverage region of a LEO satellite moves at a constant speed over the surface (approx. 7 km/s), the set of served devices changes in time.


In order to improve the revisit rate, LEO satellites are deployed in constellations.
In a \textit{dense constellation} (e.g., Starlink, Kuiper, Iridium), as one satellite hides in the horizon, another is rising to continuously serve a given device on the surface.
In these cases, Inter-satellite links (ISL) allow LEO fleets of hundreds of satellites to coordinate and relay application data with the ground station connected to the core network.
Since most IoT applications are delay-tolerant, this enables so-called \textit{sparse constellations} characterized by large coverage gaps, and opportunistic ISLs. 
This sporadic connectivity drastically reduces the fleet size requirements, to less than a dozen LEO satellites~\cite{fraire2020sparse}.
In this scenario, a new challenge appears, as data must be kept temporarily stored in satellites and/or devices until the satellite link becomes available. 

Finally, device-to-device communication~\cite{kim2018secure} can further extend the coverage of the IoT ecosystem to under-roof or underground locations leveraging repeaters and node mobility.

Next we present our recommendations on which technologies should be used in both direct and indirect satellite IoT.

\section{Technologies Enabling Satellite-IoT}\label{sec:background}

\begin{figure}[]
	\centering
	\includegraphics[width=\linewidth]{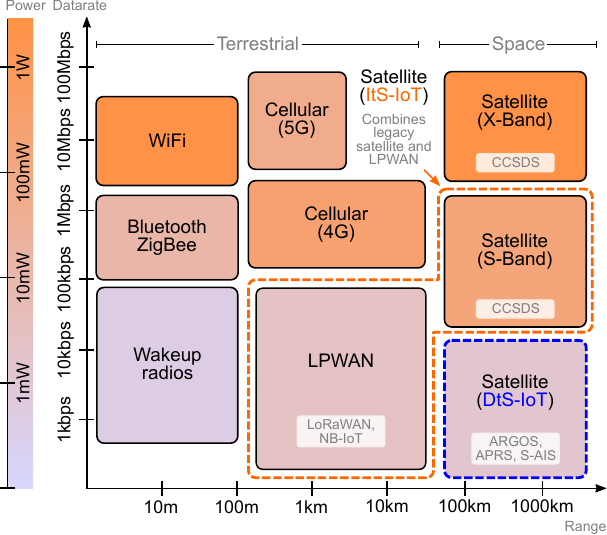}
	\caption{Protocol ecosystem in distance, data rate and power dimensions.}
	\label{fig:protocols}
\end{figure}

Fig.~\ref{fig:protocols} presents an overview of the technologies and protocols used for communication in terrestrial and space networks. 
This section presents our second contribution, which is a review and our recommendations on the technologies to be used in the \new{STEREO} architecture.

\subsection{Satellite based Communication Technologies}

Traditionally, the purpose of satellite communication was to reliably convey telecommands (TC) to the satellite, and deliver telemetry (TM) of the spacecraft to the ground station.
To this end, low data rate TM/TC protocols were standardized in the Consultative Committee for Space Data Systems (CCSDS) in the early 80s. 
S-band was (and is) the most popular TM/TC band (up to a few Mbps), over which carrier-based ranging and tracking operations also take place.
Most modern data-demanding missions are now leveraging larger bandwith over the X-Band and higher bands to convey hundreds of Mbps.
The power consumption and volume of high-throughput X-band subsystems are typically reserved for satellite platforms tens of kilograms.
CCSDS point-to-point links were also adapted for ISL applications.
Protocols like Proximity-1 or the most recent Unified Space Data Link Protocol (USLP) can operate over UHF or S-band to connect spacecraft in-orbit, although at reduced data-rates due to the long distances involved (up to a few Kbps).

\new{In the space context, multi-user application-specific low power and low data volume technologies already exist for several years, however, IoT has not been prominently addressed under such naming.
Some of the most used device-to-satellite protocols are Argos
, for environmental data (telemetry and telecontrol to weather stations and buoys), Satellite Automatic Identification System (S-AIS) 
and Satellite Automatic Dependent Surveillance–Broadcast (ADS-B)
, for tracking data from vessels and aircraft worldwide.}

\subsection{IoT Technologies}

Although a common interest, a truly global satellite IoT is more likely to be achieved if terrestrial IoT can be ported to the space domain, rather than replacing terrestrial IoT by the aforementioned space protocols.
The reasons behind this strategy are three-fold: (\textit{i}) satellite IoT could profit from mass production and derived lower costs already achieved by terrestrial IoT, specially concerning devices; (\textit{ii}) satellite IoT would benefit from a larger ecosystem and community optimizing the attainable performance with scarce resources; and (\textit{iii}) satellite IoT will seamlessly integrate with existing terrestrial deployments.
We claim that proliferated space-IoT will succeed if and only if it can profit from the economies of scale and technological advances already empowering LPWAN in terrestrial-IoT, which in turn, can significantly extend its connectivity reach to remote regions leveraging orbiting gateways

From the plethora of terrestrial IoT technologies, Low Power Wide Area Networks (LPWANs) seem the best option for blending space-terrestrial IoT technologies in a worldwide scale. 
LPWAN enable low-volume data transmission over tens of kilometers, while keeping very low energy consumption for the end-devices. 
They leverage a network architecture in which all the intelligence is moved towards a central server, allowing the development of cheap end-devices.
NB-IoT (developed by 3GPP)~\cite{sinha2017survey} and LoRa/LoRaWAN (LoRa radio combined with the LoRaWAN protocol that defines the network architecture and the communication protocols~\cite{sornin2015lorawan}) are by far the most representative ones.
Notably, recent research proved the feasibility of extending LoRa/LoRaWAN and NB-IoT with direct-to-satellite links~\cite{colavolpe2019reception,ouvry2018ultra}.

One of the key differences between these two technologies is that in NB-IoT a
given device is assumed associated and synchronized with an eNB, while LoRaWAN operates with a decoupled Aloha-based medium access protocol. 
We argue here that these technologies should both be used in the \new{STEREO} context, as needed by the application requirements. 
Indeed, NB-IoT is based on a \textit{connected} mode with a radio resource allocation and a quality of service management, thanks to the core network, that can guarantee strict Quality of Service requirements, which is of uttermost importance to some applications. 
These features comes at the expense of more complex radio access negotiations and core management elements. 
Moreover, NB-IoT as a cellular technology uses licensed frequency bands which are operated by a given telecommunication operator.
On the other hand, LoRaWAN is based on a \textit{non-connected} mode, with a low-complexity and straightforward deployment that can be of more importance to some users, rather than strict latency or reliability constraints. 
LoRaWAN uses unlicensed frequency bands which are shared among different wireless technologies. 
\new{Further background on these technologies can be found in~\cite{centenaro2021survey}.}

\subsection{\new{Satellite IoT Convergence}}

\begin{figure}
	\centering
	\includegraphics[width=\linewidth]{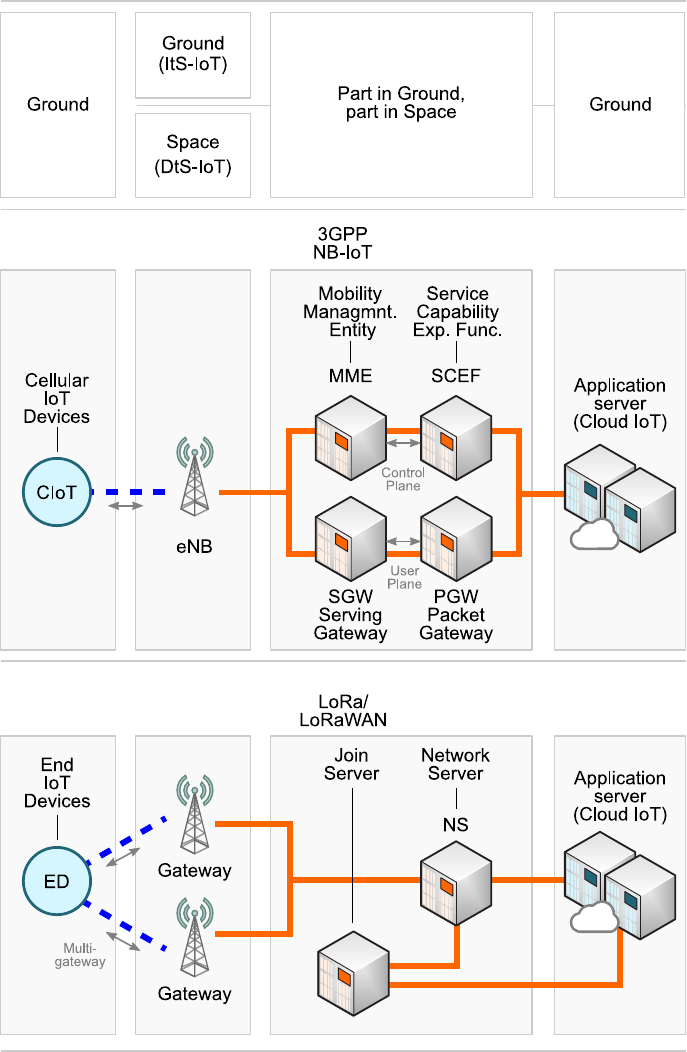}
	\caption{3GPP NB-IoT and LoRa/LoRaWAN network architectures in the Satellite IoT context.}
	\label{fig:architecture-network}
\end{figure}

Fig.~\ref{fig:architecture-network} presents a layout of our proposed NB-IoT and LoRaWAN network architecture elements in the context of \new{STEREO}. 
In all cases, IoT devices and application servers are expected to be on ground.
However, gateways/eNB can be on ground (ItS-IoT) or hosted in the satellite (DtS-IoT), or both.
As we discuss in Section~\ref{sec:challenges}, the placement of the core network elements/functionality (e.g., via virtualization) over space and ground infrastructures is still an open research question.
For instance, in a sparse IoT constellation (without immediate reach to ground), the LoRa/LoRaWAN join server should be at least partially placed in the LEO satellite to autonomously allow devices to join and relay data to the network.
\new{The main takeaway from Fig.~\ref{fig:architecture-network} is that the convergence of satellite and IoT is feasible, but brings a series of challenges regarding networking function placement and parameter optimization.} 

\section{Challenges}\label{sec:challenges}

Based on the \new{STEREO} architecture, we present here the open research challenges that will need to be addressed when IoT technologies need to work in satellite context and when satellite technologies have to handle IoT particularities. 

\paragraph{Cultural}
First of all, one of the main challenge to be addressed in satellite IoT is cultural.
Two communities from very different origins and application domain meet at \new{STEREO}.
On the one hand, a space community where trajectories and actions of a very expensive and dynamic system are known in advance, and tightly controlled by a centralized mission control.
On the other hand, a thriving terrestrial IoT sector where costs for end devices are negligible, and where control can be disregarded at the expense of achieving scalability.
Both, however, share some common aspects, such as the great importance of the scarce energy resource.

\paragraph{LEO Routing}
IoT applications have different traffic characteristics and constraints than usual satellite applications. 
In consequence, routing protocols spanning satellite-to-satellite and satellite-to-ground links considering the specifics of LoRa/LoRaWAN and NB-IoT will need to be developed for \textit{i}) dense LEO constellations with full coverage and immediate forwarding, \textit{ii}) single-satellite missions where data is collected and routed in a data mule fashion, and \textit{iii}) sparse LEO constellations with sporadic connectivity.
Route optimization criteria should be defined according to the IoT application QoS requirements (e.g., delivery delay, delivery reliability), and the concept of operations should fit the computation/memory capabilities of space assets (e.g., centralized computation on ground, distributed computation in-orbit, route table size).

\paragraph{Space Infrastructure} 
Different orbits and the heterogeneity of satellite platform families complicates a straightforward assessment of the in-orbit infrastructure.
Besides their pertinence for DtS- and ItS-IoT, the difference in LEO and GEO imposes constraints in launch and deployment costs: from just a few thousands of dollars per kilogram in LEO to 6 times more for the same mass in GEO.
Spacecrafts in GEO are also more adverse to risk than those in LEO, and the selection of flight-tested components is mandatory, which reduces the opportunities for innovation.
As a result, LEO satellites are much more attractive as test-beds for new space technology such as LoRa/LoRaWAN and NB-IoT. 
This is particularly true for CubeSats, mass and volume constrained nano-satellites (form factor of N$\times$10$\times$10$\times$10 centimeters), but highly cost-efficient thanks to a standardized mechanical and electrical framework that motivates commercial-of-the-shelf components.
The deployment of risk-prone Cube-Sat based LEO infrastructures is in contrast with legacy/conservative GEO technology; an equilibrium yet to be achieved.
All these platforms will need to co-exist in constellations defined by a set of orbital parameters that will need to be studied, analyzed and optimized from a topological perspective.

\paragraph{Mobility}
In DtS-IoT there is an opposite mobility scenario than in terrestrial networks: end-devices are mostly static, but gateways are mobile in orbit (with predictable trajectories). This opens two new challenges in IoT: (\textit{i}) how do end-devices take into account the mobility of gateways for sending their packets uplink? (\textit{ii}) how do network servers handle the scheduling of downlinks while accounting for the orbital dynamics of the satellite?
Moreover, in sparse constellations designs it is of uttermost importance to cope with disconnections (i.e., periods where there is no gateway in sight).

NB-IoT offers rich mobility features inherited from  cellular communications, including handover management and roaming. For instance, data transfers can be resumed after the device moved to another cell, thanks to mobility management involving radio resource control and core network protocols. 
In a satellite context, such an intricated core interaction will likely need to be supported by a dense constellation with persistent ISLs.
LoRaWAN, on the other hand, has a much simpler and limited mobility supported as it builds on top of a multi-gateway approach where every gateway forwards all received packets without a strict device-to-gateway bonding (thus no handover required).
It is then up to the network server to combine or discard the received frames and schedule a downlink packet when needed (e.g., an acknowledgment).
As a result, LoRaWAN, on the one side seems a good fit for sparse constellations without handover process, but on the other side, dense constellations could exploit the multi-gateway coverage, although without a strong requirement on ISLs.

In the radio access context, it remains unclear if the tighter mobility control imposed by cellular-based NB-IoT with respect to LoRaWAN can be beneficial or detrimental to \new{STEREO}.
In particular, further research is expected to understand the trade-off between energy cost and performance gain of NB-IoT and LoRaWAN mobility approaches.

\paragraph{Core Functionality Placement}
The placement of network functions in \new{STEREO} is a related and open research topic.
Both devices and satellites are resource-constrained platforms (i.e., energy, compute power, memory), thus, functions/roles must be allocated carefully.
When mounted of a satellite, the gateway or the eNB might no longer enjoy a stable and low-latency connection with the network core.
As a result, virtualization can be exploited to dynamically locate signaling functions to  decrease the control load, and to locate delay-sensitive functions at the edge of the network (i.e., the satellite), while keeping delay-tolerant aspects in the core at ground. 
For instance, ACK, HARQ, ADR, among other handshake-based features will need to be deployed and coordinated in orbit when no direct connection with the core is present.
The resulting Network Function Virtualization (NFV) 
strategy will need to cope with unprecedented connectivity gaps mandated by the predictable orbital mechanics.
Same with authentication, authorization, and accounting (AAA) functionalities, traditionally exclusively handled at the core network.
Indeed, the scalability issue highlighted in the radio access part, is also applicable the distribution of core functionalities.
On a related front, multi-gateway implications in the satellite domain remains to be explored.
For example, deciding which orbiting gateway should react with an ACK to a given message, or which one should send a beacon or downlink user data on a given instant depends on coherent but likely asynchronous core network procedures yet to be defined.

\paragraph{\new{Physical Layer}}
A core open topic in \new{STEREO} is the selection, regulation and potential licensing of a unified global-scale frequency band.
Otherwise, to support DtS-IoT technologies that work in licensed (NB-IoT) or ISM bands (LoRaWAN), the network server needs to be able to update the gateways so that the geographical area over which the LEO satellite will fly over is synchronized with the frequency and regional parameters (\textit{e.g.}, duty cycle, transmission power) defined for that area.
Also, there is room for demonstrating the performance \new{and implementation complexity of LoRa/LoRaWAN and NB-IoT transceivers} (chirp spread spectrum and narrow-band LTE modulations, respectively) in ground-space links.
\new{In fact, while satellite-to-remote regions channels do not suffer from the severe shadowing or multipath fading from urban terrestrial environments, new channel models considering the specific DtS-IoT conditions (atmosphere, Doppler, time-dynamics etc.) will need to be studied for the available physical layer configurations.}
%
\new{For example, LoRa/LoRaWAN flavors includes Sub-Gigahertz LoRa and 2.4GHz variants, as well as Frequency Hopping LoRa (LR-FHSS~\cite{boquet2021lr}), which is a modulation specifically designed to cope with the interference from large number of devices under satellite coverage}.
Also, the proper antenna design for the gateway at the satellite requires attention, \new{especially for beam-forming or MIMO multi-antenna arrays}.
All of the above could leverage existing know-how from the space community.

\paragraph{Medium Access Protocols}
On the MAC side, we need to be able to scale existing IoT protocols to handle hundreds of end devices that need to access the gateway in (the same) very short window, so that they fit the short-lived coverage of LEO satellites.
This challenge is boosted by the very long-range on the direct satellite-to-ground links with devices in DtS-IoT.
The opportunity at hand, however, is to profit from the predictable LEO satellite trajectory to determine optimal transmission spots.
Either the orbital path can be computed on constrained devices, or the satellite dynamics can be estimated by means of broadcasted parameters in beacons~\cite{vogelgesang2021uplink}, or by Doppler shift measurements on the device side.
Traffic patterns can be learned from repeating satellite passes
unlocking a proactive data aggregation and scheduling (both on device and gateway side) aiming at the optimal time to enhance the performance and reduce energy waste~\cite{ilabaca2021network}.
Thus, new access control schemes can be derived from this combined prediction, aggregation and scheduling access approaches.

\paragraph{Synchronization and Localization}
From a more general perspective, operations and network element management will rest at the core of successful \new{STEREO} systems.
On the one hand, the access part of the system will need to be enabled by access schemes that consider drifting clocks on IoT devices.
This can be achieved by MAC that either keep them synchronized, or that can operate without common time bases (i.e., LoRaWAN Class A vs. Class B).
The trade-off among both approaches is an appealing research topic.
On the other hand, the core portion of the network will need to coordinate and manage actions over asynchronous satellite-to-satellite and satellite-to-ground links.
For instance, downlink data flows could be buffered in advance in LEO satellite's memory, then scheduled  to overcome otherwise high-latency device-to-server handshakes and to save collision in such dense DtS-IoT network.
Indeed, gateway (satellite) mobility in DtS-IoT is to be managed by these same means. This overall synchronization challenge demands novel network operation concepts at the intersection of terrestrial IoT and space.
On the other hand, satellites are typically equipped with GNSS localization services, but this is not always true in constrained IoT devices.
In these cases, network-based localization solutions (e.g., leveraging beacons) will need to be created for satellite-based LoRa/LoRaWAN and NB-IoT.

\paragraph{Standardization}
To allow a seamless connectivity between ground and satellite IoT networks, the IoT stack provided by IETF should be used (UDP/CoAP/DTLS). Because of capacity limitations and delay constraints specific to the satellite context, this protocol stack should be extended, e.g., introducing a new convergence protocol for CoAP, and new multicast addresses for group communication (be it at the network or application layer). 
Also, the compression and fragmentation mechanisms provided by SCHC (RFC9011) should be adapted in this new context. 
For further standardization discussions including 3GPP group the reader is referred to~\cite{centenaro2021survey}.

\paragraph{Evaluation}
The end-to-end assessment of satellite IoT deployments, technologies and related parameters requires of new tools and metrics.
While satellite and IoT-specific simulation/emulation platforms exist, a space-terrestrial integrated solution for IoT is missing.
Only early simulator prototypes already enable first analysis spanning radio access and core network for LoRa/LoRaWAN~\cite{fraire2022florasat}.
The novelty of satellite IoT also demands the definition of new and meaningful metrics to quantify its performance.
For example, end-to-end latencies provoked by the in-orbit multi-hop forwarding needs to be closely tracked so that downlink messages fit scheduled LoRa/LoRaWAN device's reception windows.

\paragraph{Others}
Other derived aspects are relevant in future satellite IoT research.
Methodological-wise, the discipline can enjoy multiple approaches, from the optimization field, protocol design, model verification, among others.
Finally, security topics remains a relevant research topic in this context, as link intermittency in the core and access networks complicates key distribution and hinder stable encrypted data exchange.

\section{Outlook}\label{sec:conclusion}

The opportunities emerging from a global IoT vision are unprecedented, as they can impact traditionally strong business sector as well as currently unserved remote regions.
This paper has proposed a novel space-terrestrial integrated IoT architecture and has analyzed state-of-the-art space and terrestrial IoT technologies, while detecting adaptation and integration approaches. We argued that both LoRaWAN and NB-IoT are the best IoT networks for the space-terrestrial integrated Internet of Things.
Framed in this novel space IoT architecture, we were able to outline the main open research challenges laying on the path towards an IoT infrastructure with an ambitious goal of achieving a global service footprint.
In particular, we believe that the key challenges emerges from adapting the existing IoT radio access and core infrastructure to cope with the specifics of orbital dynamics: extremely long communication ranges, low capacity on board, limited energy, and frequent but predictable connectivity gaps.

\bibliography{bibfile}
\bibliographystyle{IEEEtran}

\begin{IEEEbiographynophoto}{Juan A. Fraire}
is researcher at Inria in France, and associate researcher and professor at CONICET-UNC and Saarland University in Germany. His research focuses on space networking and applications enabled by informatics techniques. Juan is the founder and chair of the STINT Workshop and has co-authored more than 70 papers. 
\end{IEEEbiographynophoto}

\begin{IEEEbiographynophoto}{Oana Iova}
is an associate professor at INSA Lyon (France). Her research is in performance evaluation, routing and MAC protocols for wireless networks, with a focus on low-power long-range technologies for the Internet of Things.  Oana is the founder and co-chair of the international NewNets Workshop and the French LPWAN Days.
\end{IEEEbiographynophoto}

\begin{IEEEbiographynophoto}{Fabrice Valois}
is full professor in INSA Lyon (France), since 2008. In 2000, he received a Ph.D. in computer science from University of Versailles (France).
Co-founder and former director of the CITI research laboratory, he is involved in the Agora Inria research team. His research interests are in the area of dynamic, dense and autonomous wireless networks.
\end{IEEEbiographynophoto}
\end{document}